%% file: main.tex
\documentclass[conference]{IEEEtran}
\usepackage{amsmath} 
\usepackage{graphicx}
\usepackage{dblfloatfix}
\usepackage{amssymb}
\usepackage[colorinlistoftodos,textwidth=3cm]{todonotes}
\usepackage{booktabs}
\usepackage{enumitem}
\usepackage{pifont}    
\usepackage{comment}
\usepackage{subcaption}
\usepackage{float}

\def\BibTeX{{\rm B\kern-.05em{\sc i\kern-.025em b}\kern-.08em
    T\kern-.1667em\lower.7ex\hbox{E}\kern-.125emX}}

\ifCLASSOPTIONcompsoc
  \usepackage[nocompress]{cite}
\else
  \usepackage{cite}
\fi
\ifCLASSINFOpdf
\else
\fi
\begin{document}
%
\title{Toward Quantum-Optimized Flow Scheduling in Multi-Beam Digital Satellites}

\author{\IEEEauthorblockN{Qiben Yan}
\IEEEauthorblockA{
\textit{Michigan State University}\\
East Lansing, Michigan 48824, USA}
\and
\IEEEauthorblockN{John P. T. Stenger}
\IEEEauthorblockA{
\textit{U.S. Naval Research Laboratory}\\
Washington, DC 20375, USA}
\and
\IEEEauthorblockN{Daniel Gunlycke}
\IEEEauthorblockA{
\textit{U.S. Naval Research Laboratory}\\
Washington, DC 20375, USA
}
}


\maketitle

\begin{abstract}
Data flow scheduling for high-throughput multi-beam satellites is a challenging NP-hard combinatorial optimization problem. 
As the problem scales, traditional methods, such as Mixed-Integer Linear Programming and heuristic schedulers, often face a trade-off between solution quality and real-time feasibility.
In this paper, we present a hybrid quantum–classical framework that improves scheduling efficiency by casting Multi-Beam Time–Frequency Slot Assignment (MB-TFSA) as a Quadratic Unconstrained Binary Optimization (QUBO) problem. We incorporate the throughput-maximization objective and operational constraints into a compact QUBO via parameter rescaling to keep the formulation tractable. To address optimization challenges in variational quantum algorithms, such as barren plateaus and rugged loss landscapes, we introduce a layer-wise training strategy that gradually increases circuit depth while iteratively refining the solution. We evaluate solution quality, runtime, and robustness on quantum hardware, and benchmark against classical and hybrid baselines using realistic, simulated satellite traffic workloads.

\end{abstract}


%
\IEEEpeerreviewmaketitle

\input{Introduction.tex}
\input{problem.tex}

\input{method.tex}

\input{evaluation.tex}
\input{discussion.tex}

\input{conclusion.tex}



\bibliographystyle{IEEEtran}
\bibliography{sat_reference}



%

\end{document}

%% file: introduction.tex
\section{Introduction}
High-throughput satellites with digital payloads employ multiple steerable beams and frequency channels to serve many users simultaneously. The rapidly-developing Low-Earth Orbit (LEO) constellations (e.g., OneWeb, SpaceX Starlink, Project Kuiper) exemplify the trend toward multi-beam satellite networks with global coverage and high capacity demands. 
Efficient on-board resource scheduling in such systems is crucial to maximize data throughput and meet quality-of-service (QoS) requirements. 

However, satellite resource scheduling faces several critical constraints. \emph{First}, there is a limited radio frequency (RF) power budget available per time slot. The satellite payload cannot activate all the beams at full power simultaneously due to power amplifier and energy storage limits. \emph{Second}, there are time-frequency slot conflicts, i.e., two transmissions cannot use the same frequency in interfering beams or the same beam at the same time without causing unacceptable interference or violating hardware limits. \emph{Third}, each user's data flow is associated with a queue of data awaiting transmission, with flow-specific priorities/deadlines. The scheduler must allocate time-frequency slots to flows in a way that respects these constraints while maximizing an overall utility (e.g., throughput). 

At its core, this scheduling task is a complex combinatorial optimization problem. It can be accurately modeled as a variant of the Generalized Assignment Problem (GAP), where transmissions (`items') must be assigned to unique resource units (`bins')~\cite{torta2025quantum}, analogous to Knapsack problems. 
Such problems are NP-hard, meaning that the search space grows exponentially with the number of flows and resources. Traditional optimization approaches based on Mixed-Integer Linear Programming (MILP) can model the problem but struggle with combinatorial explosion for realistic instances. While MILP solvers (CPLEX, OR-Tools, Gurobi, etc.) can find optimal schedules for small scenarios, they often become intractable in real-time or embedded environments (where computing resources and time are extremely limited). Even heuristic algorithms (greedy methods, genetic algorithms, etc.) face difficulty in consistently finding high-quality solutions within the time constraints. 

Recent advances in quantum computing motivate exploring alternative solution methods for such hard scheduling problems. Quantum annealers and gate-based quantum processors have shown promise on certain optimization tasks by mapping them to QUBO formulations. In particular, scheduling problems have been identified as a notable application domain for quantum optimization~\cite{kurowski2023application, guillaume2022deep}. 
Quantum  solvers (such as Ising machines or annealers) could potentially evaluate many schedule combinations in parallel via quantum superposition or specialized hardware, offering a chance to escape local optima that trap classical heuristics. Moreover, the current era of noisy intermediate-scale quantum (NISQ) devices has already enabled proof-of-concept demonstrations of solving small scheduling problems~\cite{amaro2022case}.  For example, recent work proposes a quantum approach using the ``Quantum Tree Generator" (QTG)~\cite{wilkening2025quantum}. This technique constructs a quantum state representing only the feasible solutions to knapsack-type problems, a ``hard constraint" method that avoids the complexities of penalty-based models. The QTG has been successfully adapted for the complex Quadratic Knapsack Problem (QKP) within a Quantum Approximate Optimization Algorithm (QAOA) framework~\cite{Constraint2025}, and resource analysis predicts this method could achieve a practical quantum advantage. 

In the satellite domain, NASA has explored quantum annealing for satellite mission scheduling and antenna allocation problems~\cite{biswas2017nasa}. Prior work on satellite mission planning found that classical algorithms had only incremental improvements over decades, whereas quantum methods promise more revolutionary gains in solution quality~\cite{quetschlich2023hybrid, makarov2024quantum, rainjonneau2023quantum, stollenwerk2021agile}. These early studies indicate that even though quantum hardware is not yet capable of handling industry-scale problems, hybrid approaches can already tackle medium-size instances and pave the way for future in-orbit use. 

\textbf{Our motivation} is to harness these emerging quantum techniques for the multi-beam satellite scheduling problem. The rise of NISQ devices and hybrid quantum-classical solvers (combining quantum subroutines with classical algorithms) presents a timely opportunity. If the scheduling problem can be formulated suitably for a quantum solver, we may exploit quantum resources to search the huge solution space more efficiently than a purely classical approach can. This is especially relevant for in-orbit or real-time applications, where computing time and power are at a premium. A fast, efficient quantum-enhanced scheduler could potentially be embedded onboard or used in a ground station to deliver scheduling decisions to the satellite in near real-time. Demonstrating such capability would mark a significant step toward integrating quantum computing into operational space systems. 


To anchor the research objectives, this paper poses the following key research questions in the context of satellite flow scheduling and quantum optimization:

\begin{itemize}
\item \emph{Quantum Advantage for Scheduling:} How can quantum computing methods improve the efficiency of  satellite flow scheduling under stringent resource constraints?
\item \emph{QUBO Throughput Optimization:} Can a QUBO-based formulation of the multi-beam time-frequency slot assignment problem, solved by quantum algorithms, enhance total throughput over classical methods?
\end{itemize} 

These questions aim to determine whether emerging quantum algorithms can provide a measurable benefit for complex real-time scheduling tasks aboard satellites, and how best to integrate them into existing solution frameworks. 

In solving these problems, \textbf{this paper makes the following contributions:} 

\begin{itemize}
\item We present a detailed QUBO formulation for the MB-TFSA problem that captures the primary objective of maximizing weighted throughput while encoding multiple, complex operational constraints, including per-slot power limits and per-flow queue capacities.

\item We introduce a critical data pre-processing technique, \emph{parameter rescaling}, to manage the number of required slack qubits. This method makes the QUBO model tractable for near-term quantum simulators and hardware without sacrificing the precision needed to enforce floating-point constraints, directly addressing a key barrier to practical implementation.

\item We demonstrate the necessity and effectiveness of a layer-wise training protocol for the QAOA. This strategy, which uses the optimized parameters from layer p to warm-start the search for layer p+1, proves essential for navigating the complex optimization landscape and finding high-quality solutions where standard approaches fail.


\end{itemize}

%% file: problem.tex
\section{Background and Problems}
In this section, we provide the necessary background knowledge about multi-beam satellite system and formulate the scheduling problem. 

\begin{figure}[t]
    \centering 
    \includegraphics[width=0.8\linewidth]{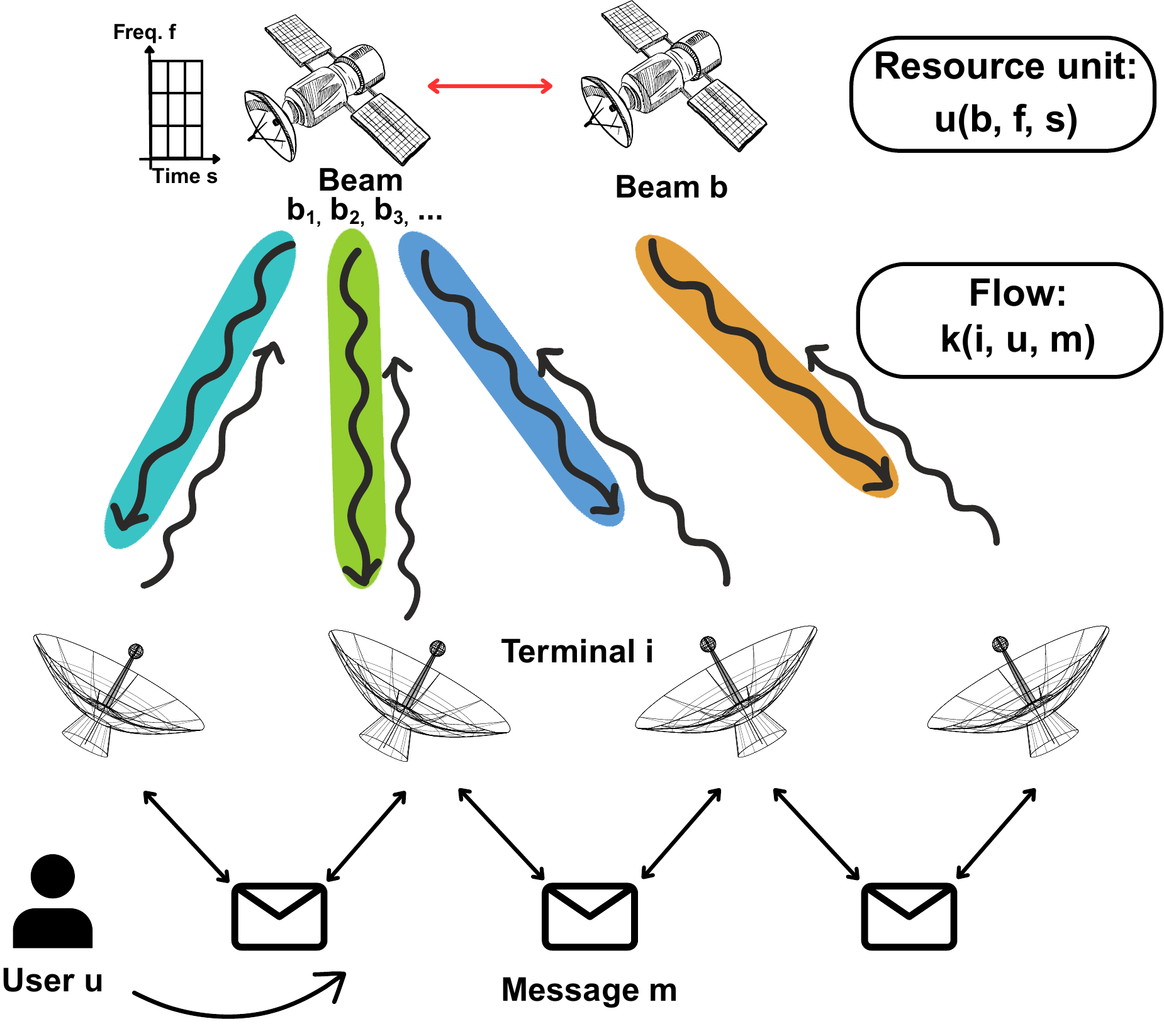} 
    \caption{Multi-beam satellite system.}
    \label{fig:system}
    \vspace{-15pt}
\end{figure}

\subsection{Multi-Beam Satellite System}

Satellite system uses multi-beam technology to increase their capacity and coverage area by simultaneously transmitting data to multiple ground locations, as shown in Figure~\ref{fig:system}. This is achieved by using phased-array antennas, which generate multiple, focused beams of radio waves, each covering a specific geographic area. A beam refers to a focused, directional stream of radio frequency (RF) energy directed toward a specific area on Earth's surface. Each beam creates a coverage cell on the ground, similar to the cells in a terrestrial cellular network. Beamforming is the technology to concentrate energy beams for intended users. We define a \textbf{resource unit}, denoted by $u$, as a unique tuple $u = u(b, f, s)$ representing a specific beam $b$, frequency channel  $f$, and time slot $s$. 


On the satellite, there are a set of data flows or service demands destined for various user terminals on the ground. Each flow can be described by attributes such as: priority weight (reflecting QoS or importance), data volume or rate requirements, current queue backlog (bits awaiting transmission), and possibly a deadline or latency tolerance. We define a \textbf{flow}, denoted by $k$, as a unidirectional stream of data packets originating from or destined for a specific user terminal. It is represented by $k = k(i, u, m)$, a tuple with a message $m$ from user $u$ on a terminal $i$.

The flow scheduling problem involves finding a  mapping of flows $k$ to resource units $u$. If frequency reuse is employed, certain flows on different beams might conflict if they use the same frequency band. This information allows us to construct constraint conditions (e.g., multiple flows cannot use the same resource units in the same slot).

\subsection{Problem Formulation}

Let us begin with a mathematical formulation of the MB-TFSA problem as a constrained optimization model. We consider a set of flows, $K$, and a set of available resource units, $U$. As previously defined, each resource unit $u\in U$ is a unique tuple $u(b, f, s)$ representing a specific beam, frequency channel, and time slot.


We first use the MILP model to capture the scheduling decisions. In a MILP formulation, we introduce binary decision variables $x_{ku}$ that equal $1$ if flow $k$ is scheduled (served) in resource unit $u$, and $0$ otherwise. The objective is to maximize the total \emph{weighted throughput}, i.e., the sum of data transmitted for each flow weighted by its priority. If $r_{ku}$ denotes the data rate (bits per slot) achievable by flow $k$ when scheduled for resource unit $u$
and $w_{k}$ is its priority weight of the flow, the objective cost can be expressed as:
\begin{equation}
\mathcal{C}_{\textrm{obj}} = -\min_{x}\sum_{k\in K}\sum_{u\in U}w_{k}\cdot r_{ku}\cdot x_{ku},
\end{equation}
summing over all flows and resource units in the scheduling horizon. This optimization is subject to multiple sets of constraints:

\noindent $\bullet$ \textbf{Resource Conflicts:} 
Each resource unit can be assigned to at most one flow. The constraint is represented as:
\begin{equation}
\sum_{k\in K}x_{ku}\leq 1, \forall u\in U.
\end{equation}

\noindent $\bullet$ \textbf{Power Budget:} The total RF power consumed by all simultaneously active beams must not exceed the available budget for each time slot. If $p_u$ is the power required for resource unit $u$, we can define $U_s$ as the set of all resource units within a given time slot $s$. The constraint for each slot $s$ is:
\begin{equation}
\sum_{u\in U_{s}}\sum_{k\in K} p_{u}\cdot x_{ku} \le P_{\max}^{(s)},
\end{equation}
where $P_{\max}^{(s)}$ is the satellite's maximum power limit for the time slot $s$, and $U_{s}$ contains all resource units in a given time slot $s$. This couples decisions across flows within the same slot. It is effectively a knapsack constraint per time slot.

\noindent $\bullet$ \textbf{Queue Capacity:} Each flow cannot send more data than it currently has in its queue.  If flow $k$ has $C_k$ bits awaiting transmission at the start of the cycle, and one slot transmission sends $r_{ku}$ bits, then: 
\begin{equation}
\sum_{u\in U_{s}} r_{ku}\cdot x_{ku} \le C_k.
\end{equation}
This ensures the schedule does not allocate more resources to a flow than needed to empty its queue (preventing wasting of resources on an empty queue).

\subsection{QUBO Transformation}

The MILP formulated above captures the problem but is difficult to solve quickly for large instances. To leverage quantum solvers, we reformulate the problem into the QUBO paradigm. In a QUBO, the goal is to minimize (or maximize) a quadratic objective function of binary variables \emph{without explicit constraints}, i.e., all constraints must be encoded as penalty terms in the objective. 

This transformation involves the following three steps: 

\noindent \textbf{1) Binary Variable Representation:} We map the scheduling decisions onto binary variables suitable for a QUBO. 

\noindent \textbf{2) Objective Term (Throughput):} The linear objective to maximize weighted throughput will be translated into a corresponding term in the QUBO objective. For a QUBO which is conventionally posed as minimization of a binary quadratic form $x^T Q x$, we can equivalently maximize throughput by \emph{minimizing} a negative weighted sum. 
This encodes the benefit of scheduling a flow transmission. It can be represented as: 
\begin{equation}
H_{\text{obj}} = -\sum_{k\in K}\sum_{u\in U}w_{k}\cdot r_{ku}\cdot x_{ku}.
\end{equation}
\noindent \textbf{3) Penalty Terms (Constraints):} All constraints are incorporated by adding quadratic penalty terms to the QUBO objective that punish any invalid combinations of variables:

\noindent $\bullet$ \textbf{Resource Conflicts:} If flows are assigned to the same resource unit, it will incur a large penalty. The QUBO form can be expressed as:
\begin{equation}
H_{1} = \sum_{u\in U}(\sum_{k \in K}x_{ku}-\frac{1}{2})^{2} - \frac{1}{4}.
\end{equation}


\vspace{3pt}
\noindent $\bullet$ \textbf{Power Budget:} This constraint ensures that no power capacity is exceeded for every time slot. This is achieved by introducing slack bits $y_{sb}$~\cite{lucas2014ising} that represents the unused power and penalizes the large fillings. It can be written as:
\begin{align}
H_{2} = \sum_{s\in S} \left[ (\sum_{u\in U_{s}}\sum_{k\in K}p_{u}\cdot x_{ku}) + (\sum_{b=0}^{M_{s}}2^{b}\cdot y_{sb}) - P_{max}^{(s)}\right]^{2},
\label{eq:slack}
\end{align}
where $M_{s} = \lfloor log_{2}P_{max}^{(s)}\rfloor$. 
The slack term, built from binary variables $y_{sb}$, converts the power limit inequality into an exact equality. These bits represent the unused capacity. When the power constraint is satisfied (i.e., consumed power  $\leq P_{max}$), the slack bits can be configured to make the penalty term ($H_{2}$) zero. 
Conversely, when the power constraint is violated, no combination of the inherently non-negative slack variables can satisfy the equality. Consequently, the penalty term will necessarily be positive, imposing a significant energy cost to the Hamiltonian for any invalid state.

\vspace{3pt}
\noindent $\bullet$ \textbf{Queue Capacity:} The QUBO of queue capacity constraint is similar to that of the power budget. We introduce slack bits $z_{kb}$ to fill the unused capacity. The QUBO form is thus written as:
\begin{align}
H_{3} = \sum_{k\in K} \left[ (\sum_{u\in U_{s}}r_{ku}\cdot x_{ku}) + (\sum_{b=0}^{M_{k}}2^{b}\cdot z_{kb}) - C_{k}\right]^{2},
\label{eq:slack2}
\end{align}
where $M_{k} = \lfloor log_{2}C_{k}\rfloor$.

After adding all necessary penalties, we obtain a QUBO objective of the form:
\begin{equation}
  H_{\text{QUBO}} = H_{\text{obj}} + \sum_{i=1}^{3} \lambda_i H_i,
\end{equation}
where the cost Hamiltonian $H_{\text{QUBO}}$ is to be minimized. A valid scheduling (all constraints satisfied) corresponds to an $x$ where no penalty is incurred, and $H_{\text{QUBO}}$ is just minus the total throughput. An infeasible or conflict-laden schedule would have positive penalty contributions, making $H_{\text{QUBO}}$ larger (worse). By choosing penalty weights $\lambda_{i}$, $i\in \{1,2,3\}$ sufficiently high, any optimal solution to the QUBO will tend to be a feasible (or near-feasible) schedule that maximizes throughput. It is important to calibrate these weights: if they are too low, the solver might prefer breaking a constraint for a slight gain in objective, yielding an invalid solution. On the other hand, if they are too high, the energy landscape becomes dominated by penalties, and the solver might struggle to differentiate among valid solutions.

\vspace{-3pt}
\subsection{Problem Size Reduction and Practical Considerations}
\vspace{-2pt}
A direct encoding of all flows, resource units, and slack bits can produce a very large QUBO matrix that exceeds current quantum hardware limits. The primary driver of this growth is the number of binary slack variables required to enforce inequality constraints with large floating-point numbers. To make the problem tractable, we employ two key techniques: \emph{slack variable quantization} and \emph{parameter rescaling}.

We introduce the parameters $dQ$ and $dP$ to control the precision of the slack variables for the queue capacity and power constraints, respectively. They introduce a fundamental trade-off. First, large $dQ (dP)$  leads to a low precision. Using a large  $dQ (dP)$ reduces the number of required slack bits, as fewer bits are needed to span the range of possible values. However, this makes the slack variable unable to precisely enforce the constraint, which can lead the optimizer to invalid or highly sub-optimal solutions. Second, small $dQ (dP)$ leads to a high precision. A small $dQ$ (e.g., $dQ=1$) allows the slack variable to represent the constraint with high fidelity. Specifically, in Eq.~(\ref{eq:slack}), the slack term $\sum_{b=0}^{M_{s}}2^{b}\cdot y_{sb}$ becomes  $\sum_{b=0}^{M_{s}}2^{b}\cdot dP \cdot y_{sb}$, where $M_{s} = \lfloor log_{2}P_{max}^{(s)}/dP\rfloor$. A similar modification is applied to Eq.~(\ref{eq:slack2}) by adding $dQ$.

However, this can lead to an explosion in the number of required qubits if the constraint's capacity ($Q$ or $P_{\text{max}}$) is large.
To overcome this trade-off, our primary strategy is aggressive parameter rescaling. We scale down all physical parameters ($r$, $Q$, $P_\text{u}$, $P_{\text{max}}$) by a common factor, which reduces their magnitude without changing the fundamental structure of the optimization problem. This allows us to use a smaller $dQ$ and $dP$ while keeping the total qubit count manageable.

The number of slack qubits required for an inequality constraint is determined by the ratio of the capacity $Q$, to the parameter $dQ$. A low ratio is necessary to maintain a manageable qubit count for near-term hardware and simulators. For instance, consider a flow with a queue capacity of $Q=1,000$. A low-precision approach using a large $dQ$ (e.g., $dQ=600$) results in a low ratio ($Q/dQ\approx 1.7$) and few slack bits, but it cannot accurately model the constraint when data rates have a finer granularity. Conversely, a high-precision approach ($dQ=1$) provides a high-fidelity discretization of the constraint but results in a large ratio ($Q/dQ=1,000$) and a big number of slack bits ($10$ in this case).

Our strategy of aggressive rescaling addresses this trade-off directly. By applying a scaling factor of $500$, we reduce the effective capacity to $Q_{\text{scaled}}=2.0$, and can set $dQ=1.0$, which maintains the desired low ratio ($Q_{\text{scaled}}/dQ=2.0$) and requires only 2 slack bits. 
We choose this scaling so that $Q$ matches the smallest rate increment that is meaningful in our setting (and is within the noise/variability of the data rates), while keeping the qubit count computationally feasible. If finer resolution is needed, a smaller scaling factor (or smaller $dQ$) can be used at the cost of additional slack bits and qubits.

Through these methods, we aim to generate a QUBO with fewer binary variables, which is within the tractable range of current quantum hardware and shallow QAOA circuits on a few dozen qubit gate-model machine. 
It is worth noting that formulating the problem as a QUBO is not lossless, as it transforms a constrained problem into an unconstrained one with penalties, which means if the solver is not optimal, it might return solutions that violate some constraints slightly (e.g., low-penalty conflicts). We mitigate this by appropriate penalty tuning and by a post-processing check: any solution from the quantum solver can be verified for feasibility and repaired by a local heuristic if needed (for instance, if two conflicting flows were scheduled, drop the lower priority one). Empirically, strong penalty weights and good initialization (discussed in the next subsection) keep the solver in the feasible region predominantly. 


%% file: method.tex
\section{Methods}
Having formulated the scheduling problem as a QUBO, we now describe the solvers, including classical baseline methods and the proposed hybrid quantum-classical algorithms.

\subsection{Classical Baseline Solver}
As a benchmark, we use a classical optimization approach to solve the scheduling problem. Where possible, we solve the MILP formulation exactly using a state-of-the-art solver (such as Solving Constraint Integer Programs (SCIP)~\cite{achterberg2009scip}). The solvers employ branch-and-bound and cutting-plane techniques and can often find optimal or near-optimal solutions for moderate-sized instances given enough time. The MILP optimum serves as an upper bound on achievable weighted throughput. However, for larger instances where MILP cannot be solved to optimality within a practical time limit, we use two alternatives: (a) a heuristic algorithm (e.g., a greedy slot allocator) that is designed for fast decisions, and (b) the best solution the MILP solver can find in a limited runtime. The classical solver yields a baseline throughput and records if any constraints were violated in its solution (ideally none). It also provides a baseline computation time and resource usage. 

\subsection{Quantum Solvers}
To generate the QUBO instance from a given scheduling scenario, we implement a QUBO construction module.  Once the QUBO is constructed, we design multiple solver options. 


\begin{figure}[t]
    \centering 
    \includegraphics[width=\linewidth]{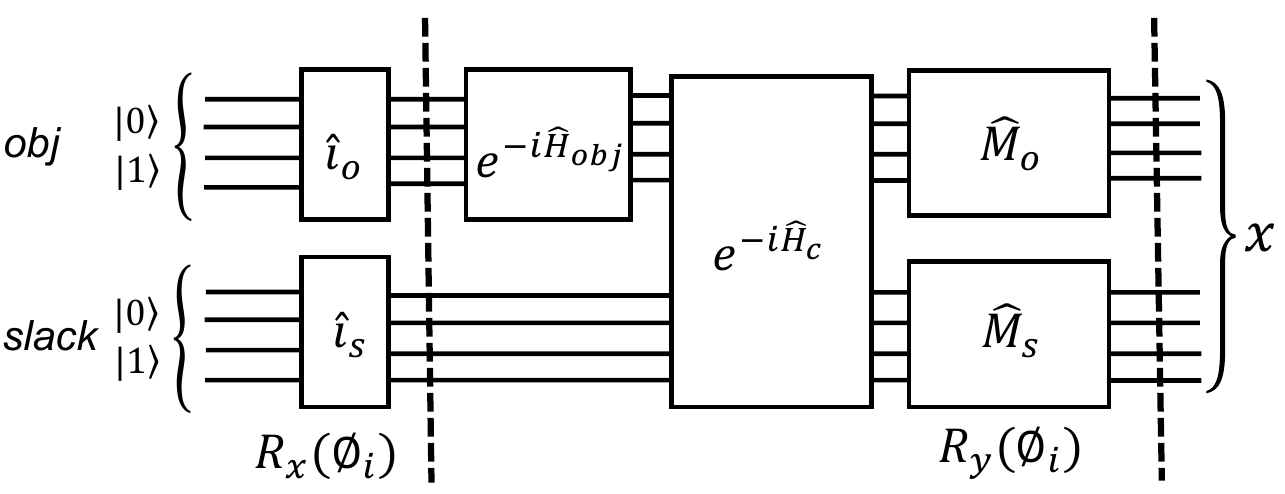} 
    \caption{One-layer QAOA solver overview (the components in between the dashed line can be replicated to add more layers). }\label{fig:qaoa}
    \vspace{-15pt}
\end{figure}

\textbf{QAOA Solver:} We experiment with gate-model quantum computers via the Quantum Approximate Optimization Algorithm (QAOA)~\cite{farhi2014quantum}. QAOA is a variational quantum algorithm well-suited for QUBO/Ising problems. In QAOA, we prepare a parameterized quantum circuit consisting of alternating layers of problem Hamiltonian evolution and mixing operations. The parameters (i.e., angles) are optimized by a classical optimizer to minimize the QUBO objective expectation value. We implement QAOA and verify its implementation on IBM quantum devices or simulators. Due to limited qubit counts on gate machines (tens of qubits), we target smaller instances. We consider low-depth QAOA (depth $p=1$ to $3$) because deeper circuits are challenging on noisy hardware. We execute QAOA multiple times from random initial parameter seeds to search the parameter landscape. 

Figure~\ref{fig:qaoa} shows the structure of the one-layer QAOA solver. QAOA prepares the ground state of the problem Hamiltonian $H_{QUBO}$ by iteratively applying a sequence of unitary operators to an initial state. The system state is defined over two primary qubit registers: the objective register $|obj\rangle$, representing the decision variables, and the slack register $|slack\rangle$, used to handle inequality constraints. The circuit begins with initial state preparation unitaries $\hat{l}_o$ and $\hat{l}_s$, followed by a phase separation stage. This stage consists of the objective Hamiltonian evolution $e^{-i\hat{H}_{obj}}$ and a joint constraint Hamiltonian evolution $e^{-i\hat{H}_{c}}$ that couples the objective and slack registers. The mixing stage is defined by the operators $\hat{M}_o$ and $\hat{M}_s$ for the respective registers. The variational evolution is governed by the parameter sets $R_x(\phi_i)$ and $R_y(\phi_i)$, representing the rotation angles optimized by the classical feedback loop. The final output $x$ is obtained by measuring the registers in the computational basis, yielding a candidate solution to the optimization problem.
It follows the following steps: 

\noindent \textbf{1. Initial State:}  The algorithm begins in an equal superposition of all possible computational basis states. This state is prepared by applying a Hadamard gate ($\mathcal{H}$) to each of the $N$ qubits, which are initialized to the $|0\rangle$ state:
\begin{equation}
|\psi_0\rangle = \mathcal{H}^{\otimes N} |0\rangle^{\otimes N} = \frac{1}{\sqrt{2^N}} \sum_{z=0}^{2^N-1} |z\rangle.
\end{equation}

\noindent \textbf{2. Variational Ansatz:}  The core of QAOA is the variational circuit, or ansatz, which is composed of p layers. Each layer consists of two unitary operators parameterized by angles $\gamma_{i}, \beta_{i}$. First, the Cost Operator, $U_C(\gamma_i)$,  applies the problem Hamiltonian $H_{\text{QUBO}}$
for a duration $\gamma_{i}$. It encodes the cost function by imparting a phase to each computational basis state proportional to its cost:
\begin{equation}
U_C(\gamma_i) = e^{-i\gamma_i H_{\textrm{QUBO}}}.
\end{equation}
Second, the Mixer Operator $U_B(\beta_i)$ applies a mixer Hamiltonian, $H_B$, for a duration $\beta_{i}$. The standard mixer is a transverse field applied to all qubits, which induces transitions between different basis states, allowing the algorithm to explore the solution space.
 \begin{equation}
 H_B = \sum_{j=1}^{N} X_j \quad \implies \quad U_B(\beta_i) = e^{-i\beta_i H_B} = \prod_{j=1}^{N} e^{-i\beta_i X_j}.
 \end{equation}
The final state after $p$ layers is given by:
 \begin{equation}
|\psi_p(\boldsymbol{\gamma}, \boldsymbol{\beta})\rangle = U_B(\beta_p) U_C(\gamma_p) \cdots U_B(\beta_1) U_C(\gamma_1) |\psi_0\rangle.
 \end{equation}

\noindent \textbf{3. Classical Optimizer:} A classical optimizer (in our case, the Simultaneous Perturbation Stochastic Approximation (SPSA) algorithm) tunes the $2p$ variational parameters $(\gamma,\beta)$ to minimize the expectation value of the problem Hamiltonian, which represents the final cost:
\begin{equation}
\min_{\boldsymbol{\gamma}, \boldsymbol{\beta}} \langle \psi_p(\boldsymbol{\gamma}, \boldsymbol{\beta}) | H_{\textrm{QUBO}} | \psi_p(\boldsymbol{\gamma}, \boldsymbol{\beta}) \rangle.
\end{equation}
 The final optimized parameters produce a quantum state from which the best solution candidates are determined by measurement.

\textbf{Alternative Variational Ansatz with Custom $R_{y}$ Mixer:} In addition to the standard QAOA, we explored an alternative ansatz with a more complex, state-dependent mixer. This approach differs from the standard QAOA in two key ways. First, instead of a uniform superposition, the initial state is prepared by applying a parameterized Ry rotation to each qubit:
\begin{equation}
|\psi_0(\boldsymbol{\phi})\rangle = \bigotimes_{j=1}^{N} R_y(\phi_j) |0\rangle,
\end{equation}
where the angles $\phi$ are fixed hyperparameters, typically derived from a classical solution where $\phi_{j}\in \{0, \pi\}$. 

Second, the standard mixer is replaced with a custom mixer that is associated with the initial state angles $\phi$. The unitary for this mixer is:
\begin{equation}
    U_{B'}(\beta_i, \boldsymbol{\phi}) = \prod_{j=1}^{N} \left[ R_y(\phi_j) R_z(\beta_i) R_y(-\phi_j) \right]. 
\end{equation}
This mixer is designed to perform a ``local" search, exploring states in the vicinity of the initial state defined by $\phi$. This mixer is chosen so that $|\psi_0(\boldsymbol{\phi})\rangle$ is an eigenvector of $H_{B'}(\boldsymbol{\phi})$ where $U_{B'}(\beta_i,\boldsymbol{\phi}) = e^{i H_{B'}(\boldsymbol{\phi)} \beta_i}$.

\begin{figure*}[t]
    \centering
    \begin{subfigure}[t]{0.32\textwidth}
        \centering
        \includegraphics[width=\linewidth]{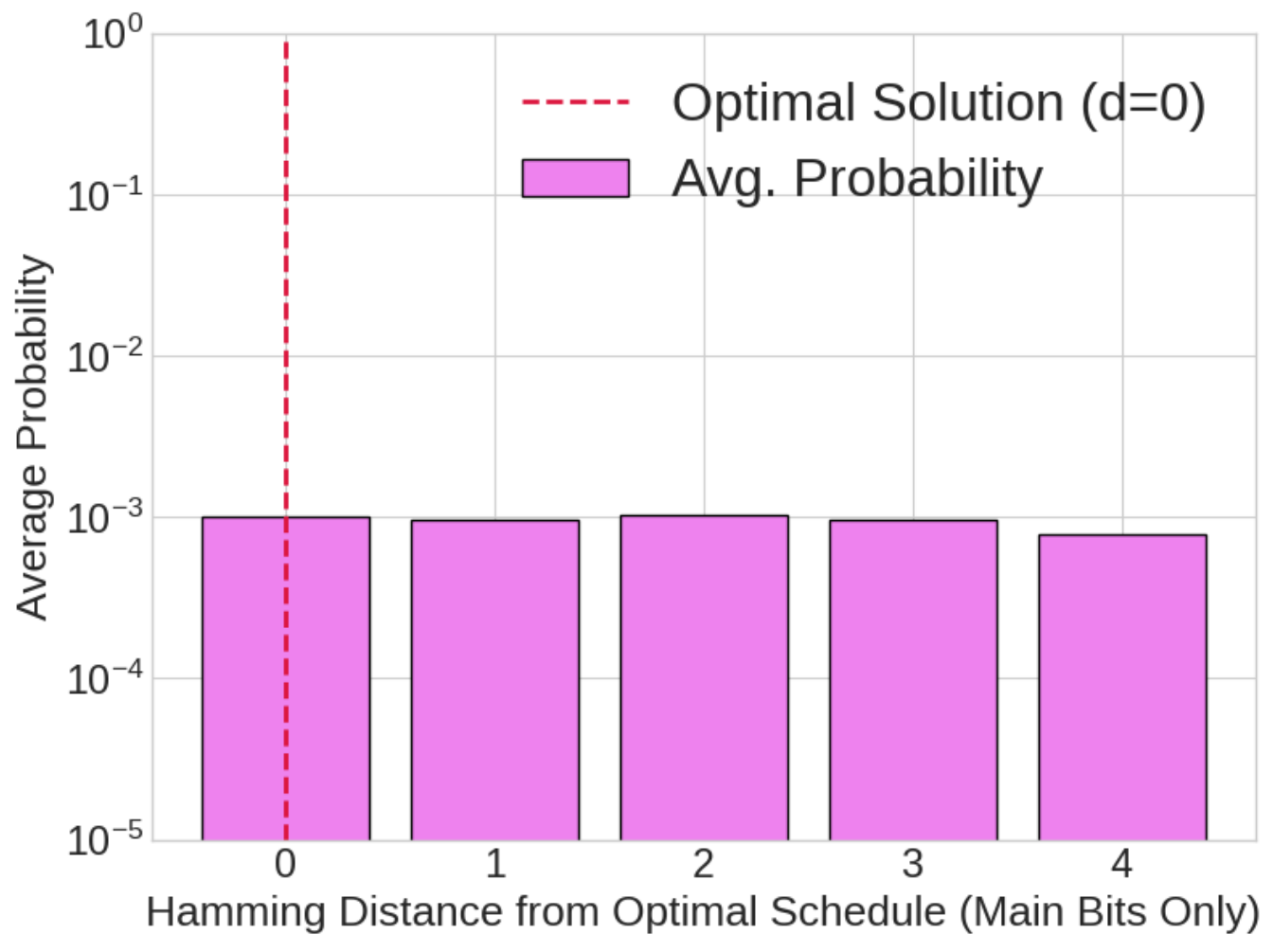}
        \caption{Solution distribution for 2-flow problem}
        \label{fig:prob-hamming-flow2}
    \end{subfigure}
    \hfill
    \begin{subfigure}[t]{0.32\textwidth}
        \centering
        \includegraphics[width=\linewidth]{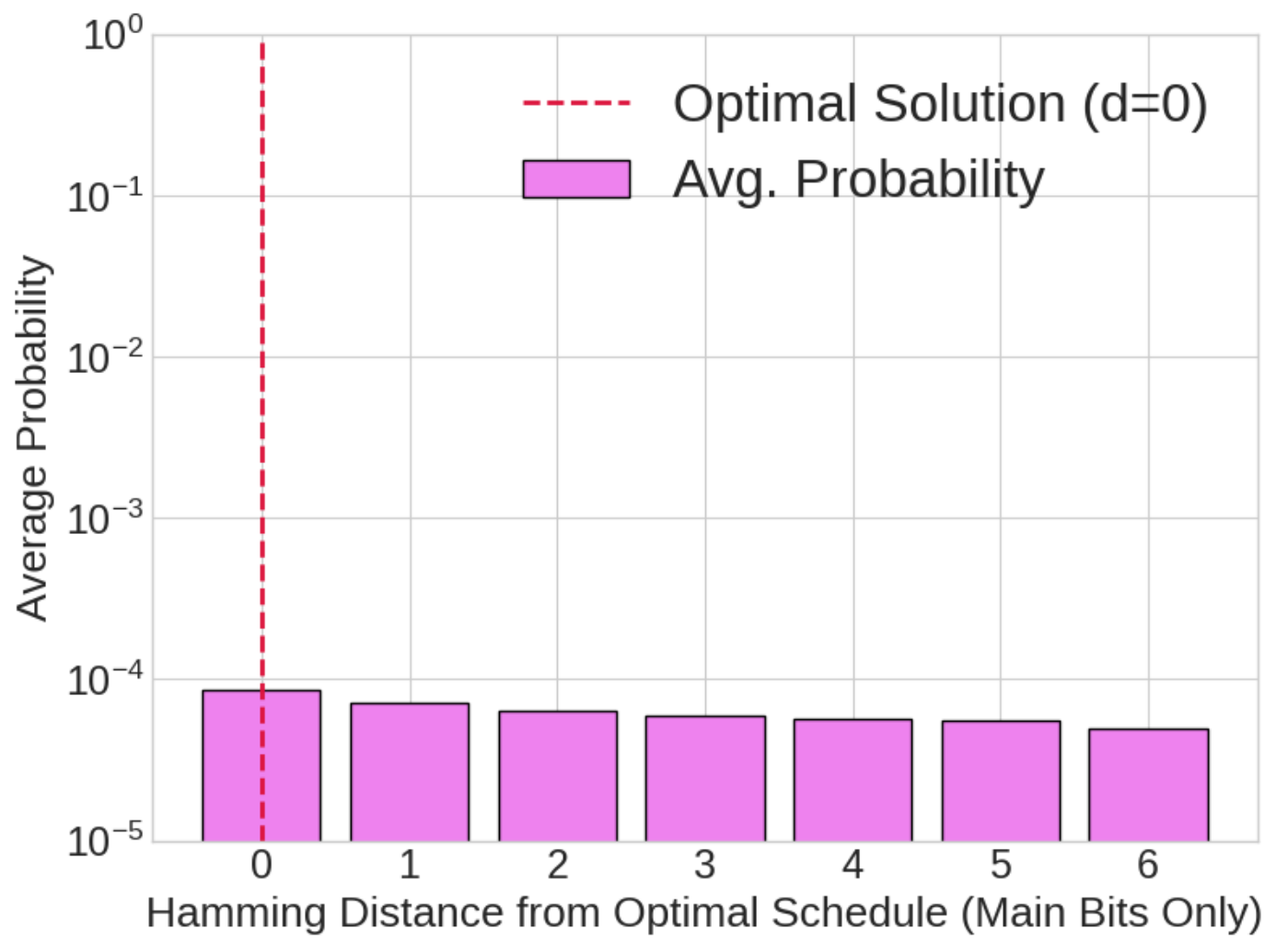}
        \caption{Solution distribution for 3-flow problem}
        \label{fig:prob-hamming-flow3}
    \end{subfigure}
    \hfill
    \begin{subfigure}[t]{0.32\textwidth}
        \centering
        \includegraphics[width=\linewidth]{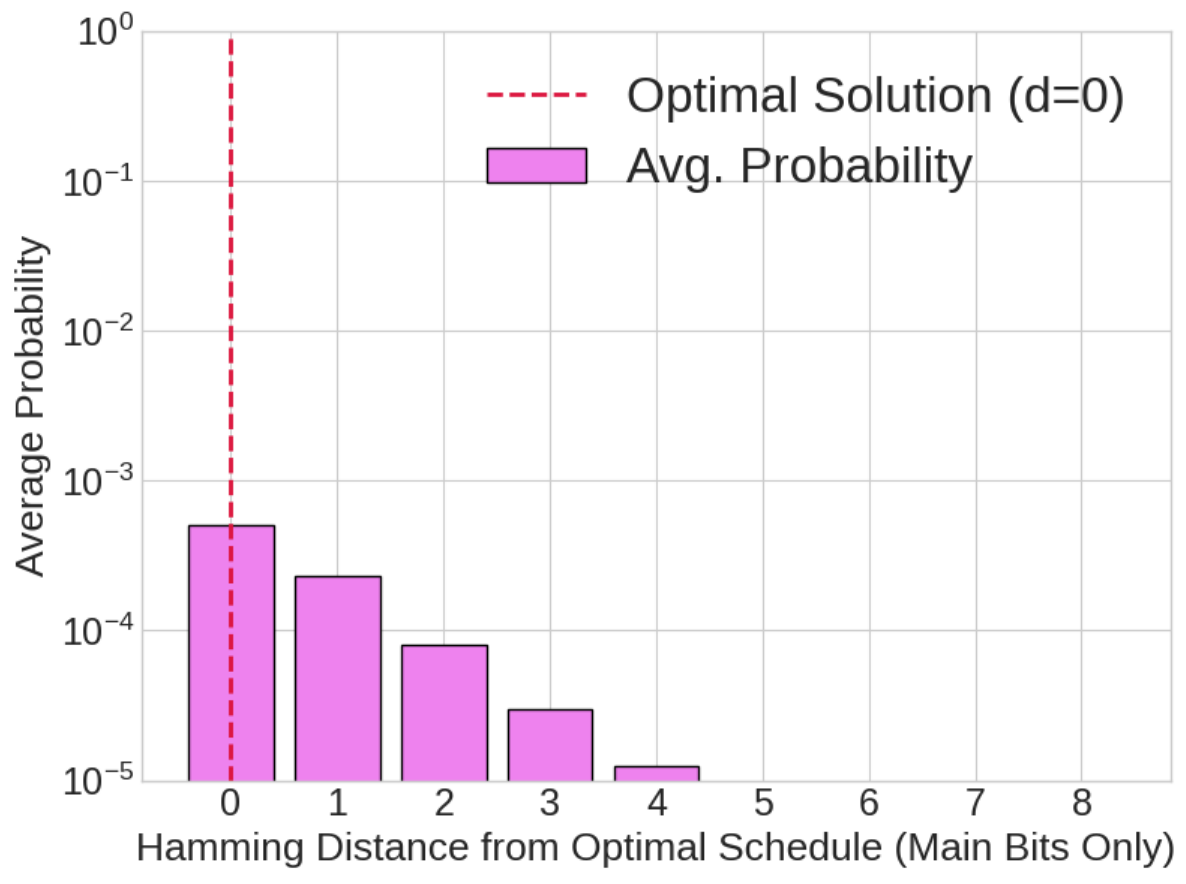}
        \caption{Solution distribution for 4-flow problem}
        \label{fig:prob-hamming-flow4}
    \end{subfigure}
    \caption{QAOA solution distribution vs. hamming distance.}
    \vspace{-15pt}
    \label{fig:prob-hamming}
\end{figure*}

\subsection{Layer-wise QAOA}

A standard QAOA with a deep circuit (large p) presents a formidable challenge for classical optimizers due to the high-dimensional and often rugged nature of the parameter landscape. To address this, we employ an iterative, layer-wise training strategy, 
\emph{Layer-wise QAOA}, which gradually increases the circuit depth: start with $p=1$ from the warm start, optimize, then use the result as initialization for $p=2$, and so on.  This method builds the complexity of the ansatz gradually, which has been shown to help the optimizer converge to better solutions more reliably than a single, ``one-shot" optimization.

The procedure is as follows:

\begin{enumerate}
   \item \textbf{Initialize:} Begin with a simple p=1 QAOA, with randomly initialized parameters ($\gamma_{1}$, $\beta_{1}$). 
   \item \textbf{Optimize:} Run the classical optimizer (SPSA) to find the optimal parameters for the p=1 circuit.
   \item \textbf{Iterate and Extend:} Use the optimized parameters from the p-layer circuit as a ``warm start" for the (p+1)-layer circuit. The first p parameters are fixed to their previously found optimal values, and the optimizer's task is to find the best values for the new parameters ($\gamma_{p+1}$, $\beta_{p+1}$).
   \item \textbf{Repeat:} Continue this process until a desired maximum depth $p_{\text{max}}$ is reached.
\end{enumerate}

This incremental layering guides the optimizer through a sequence of simpler, lower-dimensional problems, preventing it from getting stuck in poor local minima. It provides a robust path toward a high-quality solution.

\section{Data Sets}
For empirical evaluation, we test our algorithms on a range of simulated satellite scheduling scenarios. The data sources and generation methods are as follows:

\textbf{Simulated Multi-Beam Satellite Data:} We simulate a generic high-throughput satellite with multiple beams covering different cells on Earth. We assign a power budget per resource unit. These system parameters are inspired by real systems, for instance, O3b MEO satellites have on the order of 10 beams and specific power constraints~\cite{torkzaban2023capacitated}.

\textbf{Traffic Demand Profiles:} We create different traffic distributions to test robustness. For example: \emph{Uniform random flows:} a number of flows randomly assigned to beams such that each beam has similar load. \emph{Skewed (hotspot) traffic:} a scenario where one or two beams have a disproportionate number of high-priority flows (to simulate hotspots or beams covering busy areas), while others are lightly loaded. \emph{Mixed priority:} flows have a mix of priority values (some critical, some low) to test how well algorithms prioritize. 
For each flow, we assign a random data volume, and a priority weight (which could be correlated with volume or independent).

\section{Evaluation Criteria}


We first evaluate how good the schedules produced by each method are in terms of the primary objective (throughput) and how close they are to optimal.

\textbf{Weighted Throughput Achieved:} This is the objective value of the solution, $\sum_{ku} w_{k} r_{ku} x_{ku}$, which represents the total priority-weighted bits served in the scheduling horizon. We will compare the weighted throughput of different solvers. The MILP optimum will be considered 100\% and other methods will be expressed relative to this. Higher throughput is better.


\textbf{Hamming Distance to Optimum:} The Hamming distance measures the structural difference between the solution bitstring found by a solver and the true optimal bitstring. It is calculated as the number of bit positions at which the two strings differ. A lower Hamming distance indicates that the found solution is structurally more similar to the optimal one. A distance of 0 means the exact optimal solution was found. This metric provides insight into how close an algorithm gets to the correct assignment configuration.

%% file: evaluation.tex
\begin{figure*}[t]
    \centering
    \begin{subfigure}[t]{0.32\textwidth}
        \centering
        \includegraphics[width=\linewidth]{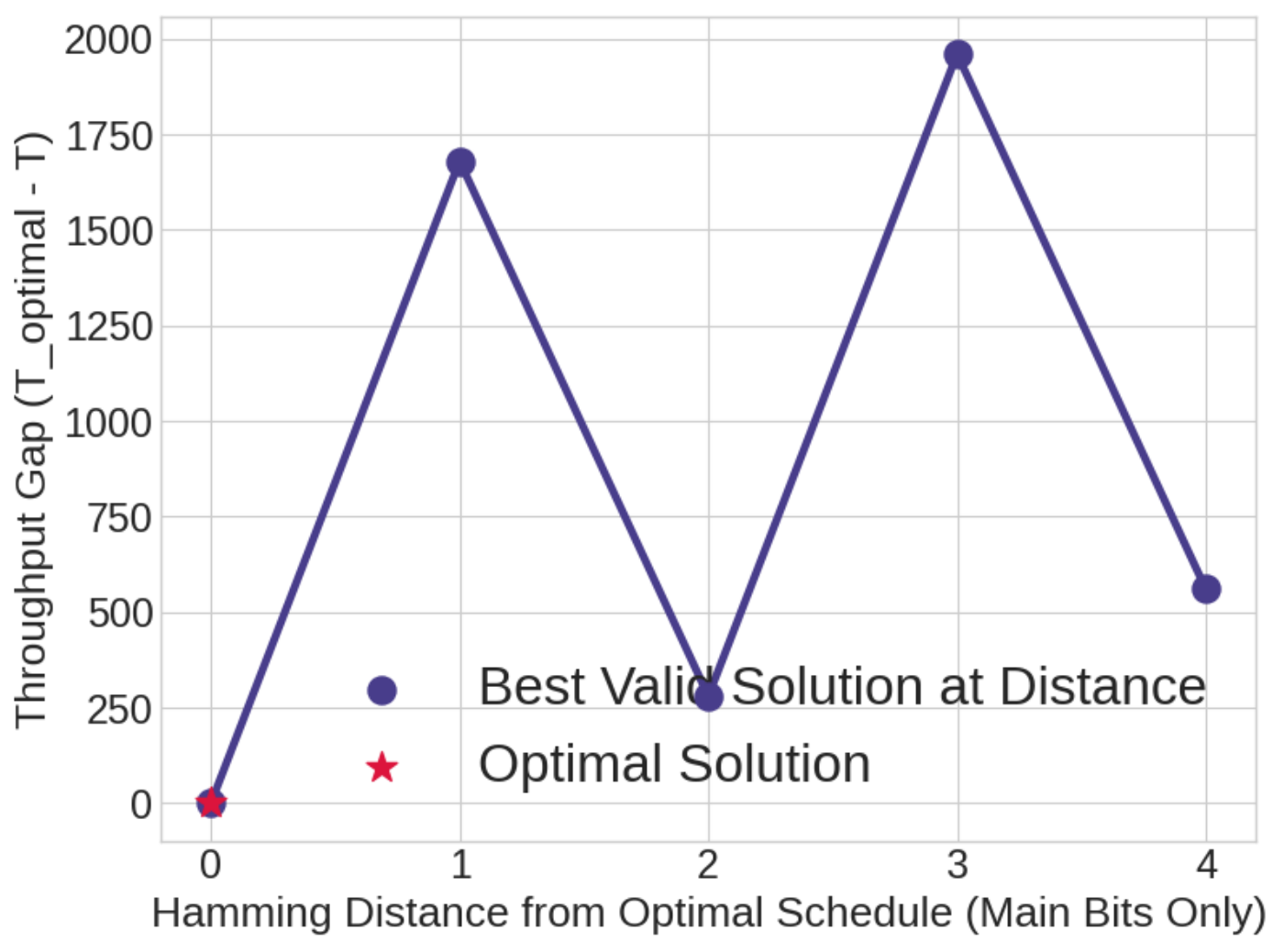}
        \caption{2-flow problem}
        \label{fig:throughput-hamming-flow2}
    \end{subfigure}
    \hfill
    \begin{subfigure}[t]{0.32\textwidth}
        \centering
        \includegraphics[width=\linewidth]{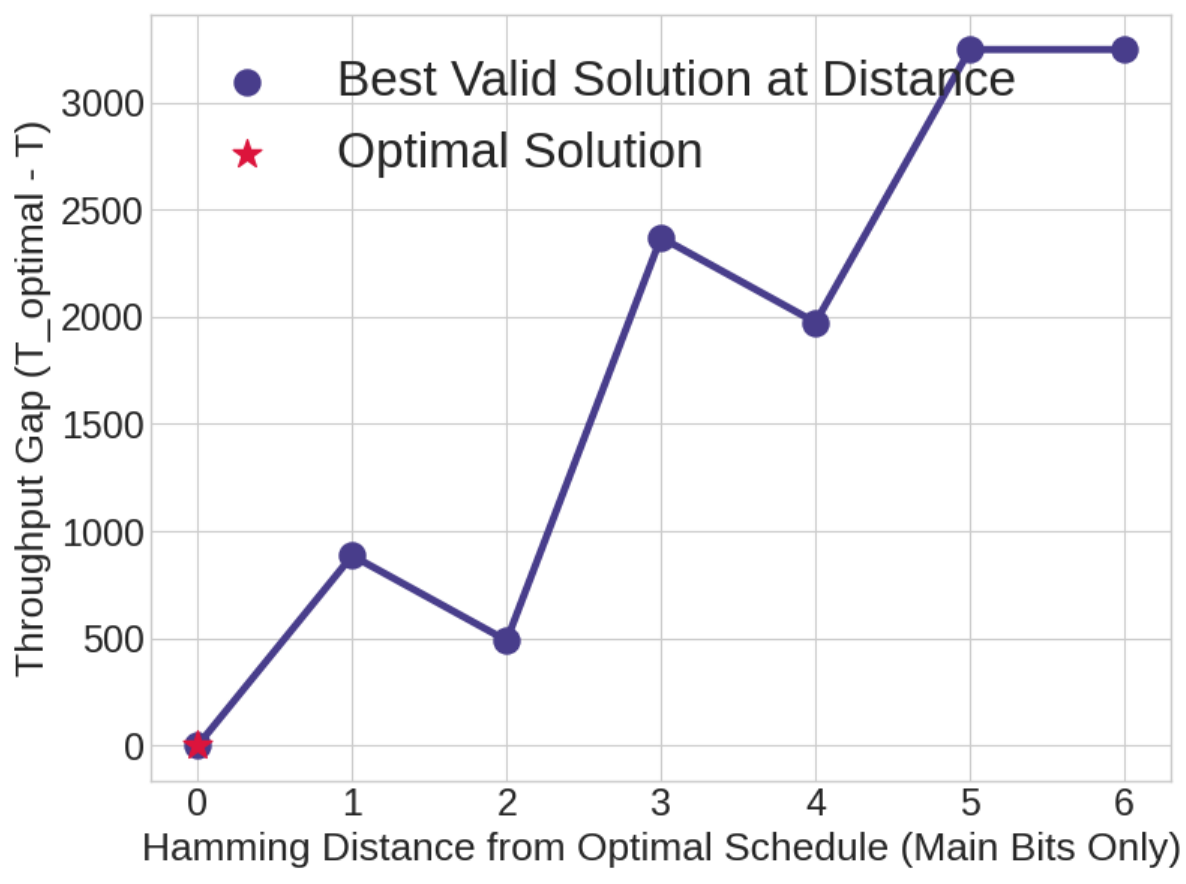}
        \caption{3-flow problem}
        \label{fig:throughput-hamming-flow3}
    \end{subfigure}
    \hfill
    \begin{subfigure}[t]{0.32\textwidth}
        \centering
        \includegraphics[width=\linewidth]{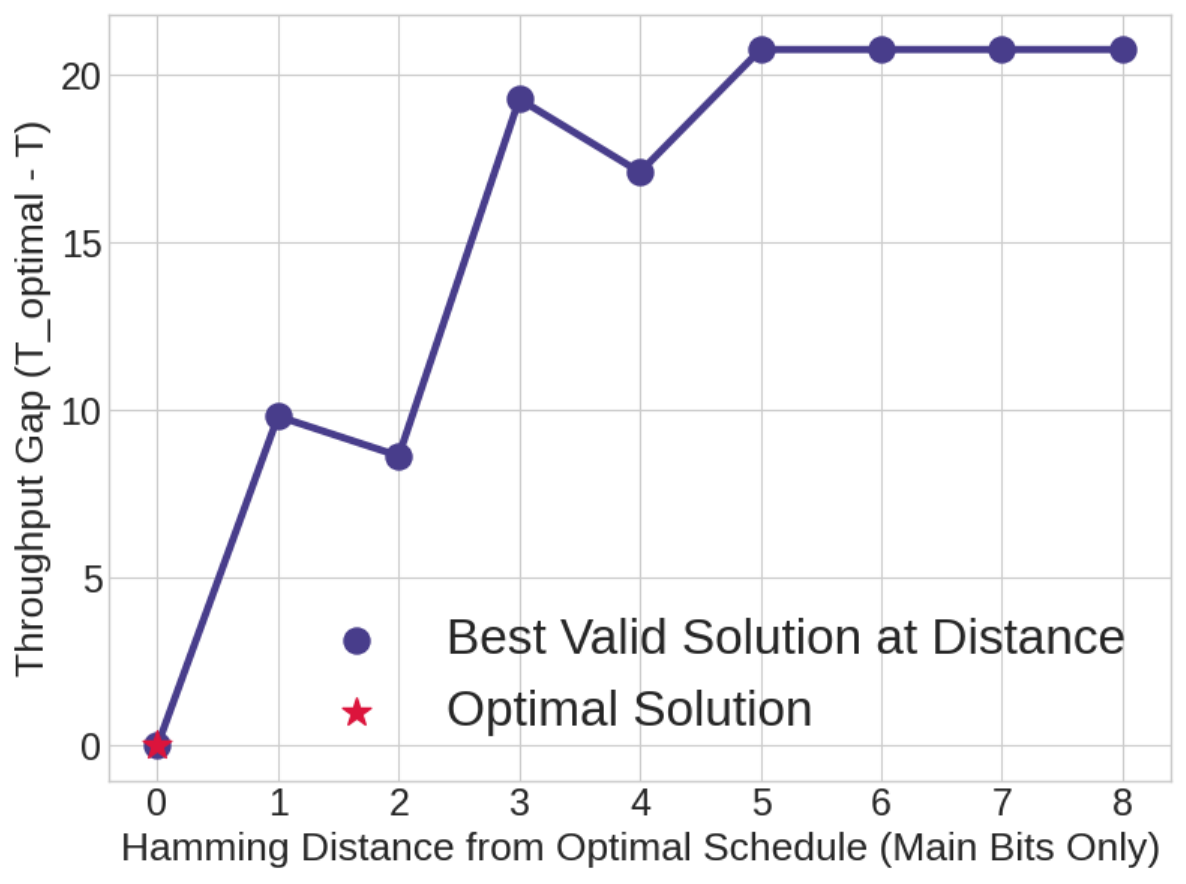}
        \caption{4-flow problem}
        \label{fig:throughput-hamming-flow4}
    \end{subfigure}
    \caption{Throughput vs. hamming distance.}
    \vspace{-15pt}
    \label{fig:throughput-hamming}
\end{figure*}

\section{Results} \label{sec:results}

In this section, we present the empirical results of our investigation into solving the MB-TFSA problem using the proposed QAOA framework. We analyze the performance of the standard p=1 QAOA on problems of increasing complexity, evaluate the impact of a greedy warm-starting technique, discuss the challenges encountered when increasing the depth of the quantum circuit, and assess the robustness of the proposed methods. We implement the algorithms using \emph{Python 3.13.5} and \emph{Qiskit 1.4.2}. All experiments were conducted on the \textbf{IBM Quantum Cloud platform}, with each optimization step utilizing 4,096 shots to evaluate the expectation value. Particularly, we use \emph{ibm\_torino} 
for running all the experiments.

\subsection{Performance of 1-Layer QAOA}
We apply a standard, 1-layer (p=1) QAOA to three experimental setups of increasing complexity: a 2-flow, 3-flow, and 4-flow scheduling problem, each with two available resource units. We note that the model could assign the same flow to multiple resource units provided the queue capacity is not exceeded.


Our key finding reveals a non-monotonic relationship between problem size and solution difficulty. The p=1 QAOA successfully identifies the optimal throughput for the 2-flow and 4-flow problems. In contrast, for the 3-flow instance, the algorithm settles in a sub-optimal local minimum. However, we found that this sub-optimal solution contained the structure of the true optimum, which could be recovered via a simple post-processing step to resolve a constraint conflict. This indicates that while the p=1 ansatz identified the correct high-value assignments, it failed to resolve their interaction correctly, which is a characteristic failure mode for shallow circuits on complex energy landscapes.


Further insight is provided by the distribution of solutions relative to the optimal schedule, shown in Figure~\ref{fig:prob-hamming}. For the 4-flow problem, we observe a desirable, sharp decay in probability as the Hamming distance from the optimal solution increases. This indicates a high-fidelity final state, well-localized around the true optimum. The distributions for the 2-flow and 3-flow problems are flatter, suggesting that the final state is more diffuse across the solution space. The plot of the throughput gap vs. Hamming distance (Figure~\ref{fig:throughput-hamming}) visualizes the problem's cost landscape. While there is a general trend of a larger throughput gap for solutions further from the optimum, the path is not monotonic, which indicates the existence of many local optima that can trap a heuristic solver.

\begin{figure*}[t]
    \centering
    \begin{subfigure}[t]{0.32\textwidth}
        \centering
        \includegraphics[width=\linewidth]{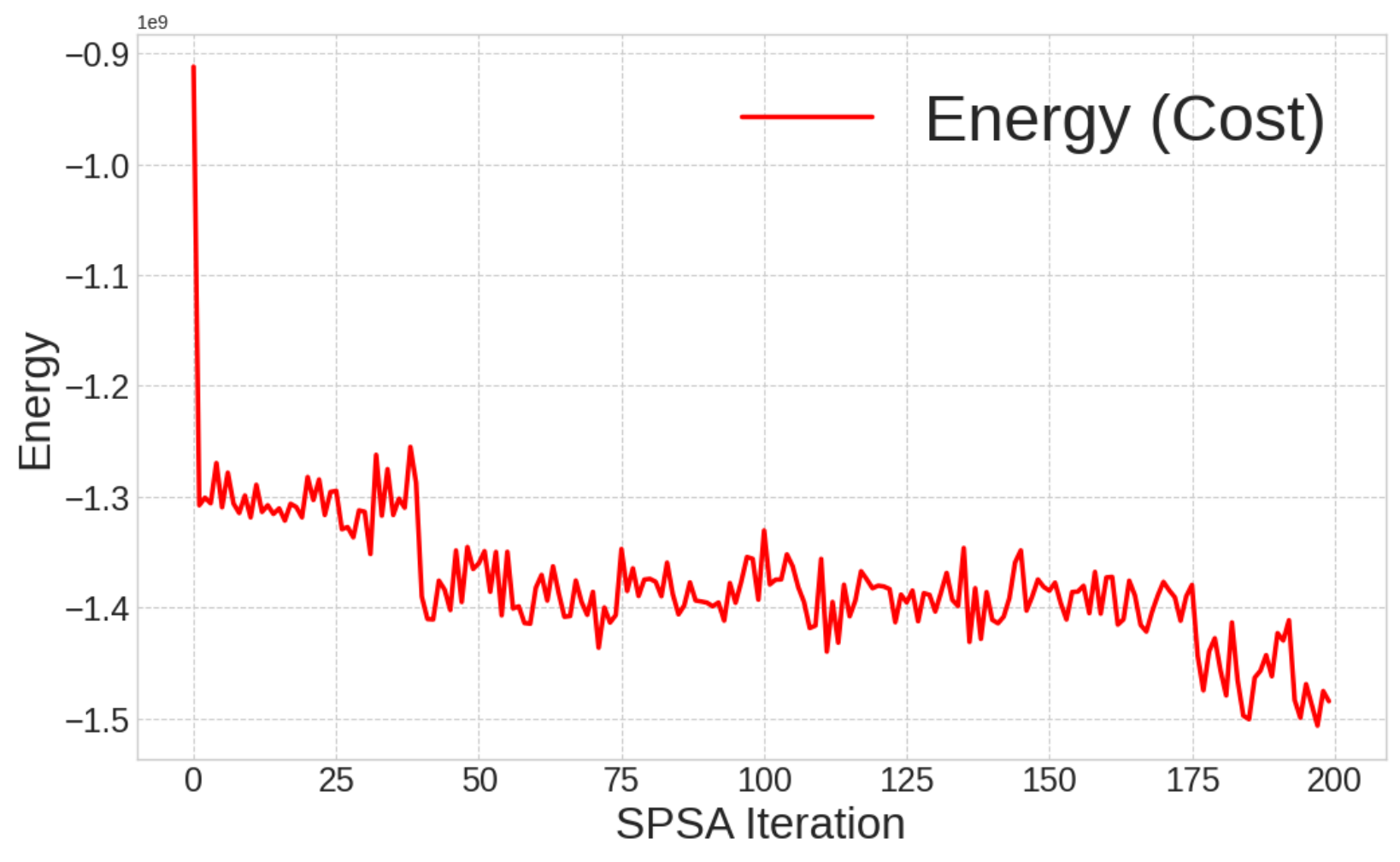}
        \caption{Energy of optimization for 2-flow setup}
        \label{fig:energy-flow2}
    \end{subfigure}
    \hfill
    \begin{subfigure}[t]{0.32\textwidth}
        \centering
        \includegraphics[width=\linewidth]{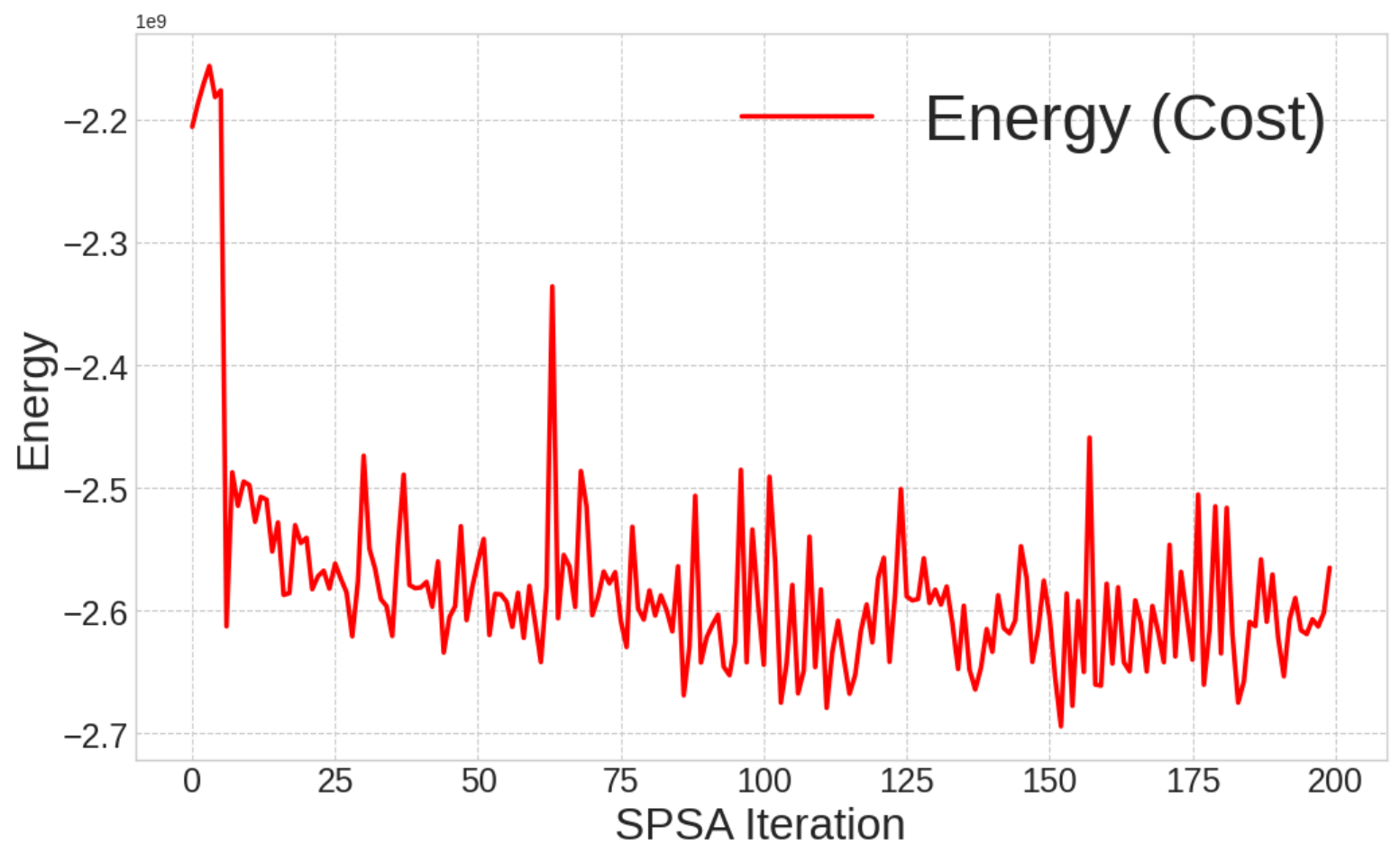}
        \caption{Energy of optimization for 3-flow setup}
        \label{fig:energy-flow3}
    \end{subfigure}
    \hfill
    \begin{subfigure}[t]{0.32\textwidth}
        \centering
        \includegraphics[width=\linewidth]{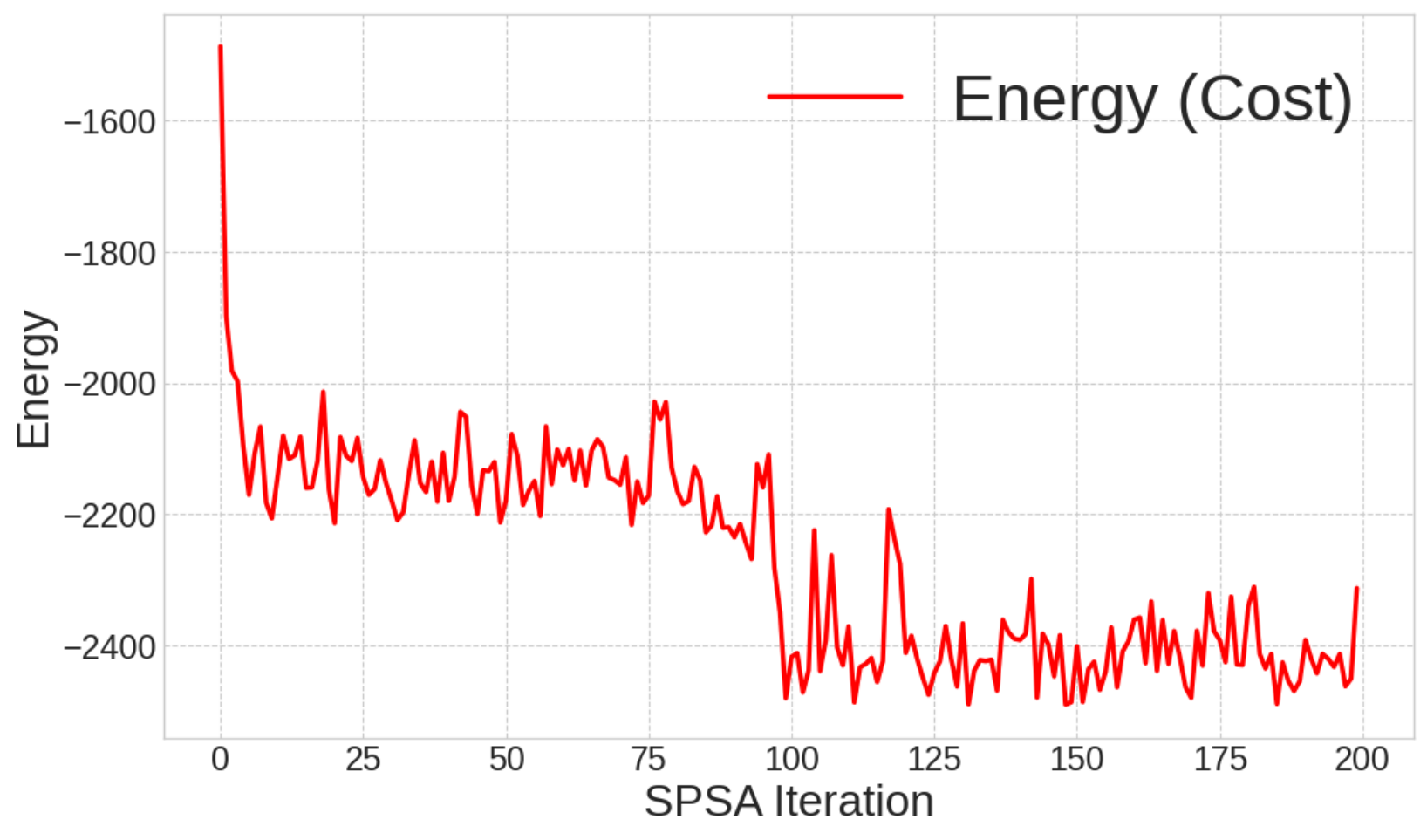}
        \caption{Energy of optimization for 4-flow setup}
        \label{fig:energy-flow4}
    \end{subfigure}
    \caption{Energy vs. SPSA iterations.}
    \vspace{-15pt}
    \label{fig:energy-flow}
\end{figure*}


\vspace{-3pt}
\subsection{Energy Convergence Analysis}
\vspace{-5pt}
It is worth noting that our implementation of the SPSA optimizer utilizes a random-search hill-climbing update for the parameters $\phi$, rather than a conventional gradient-based descent. In this approach, a perturbed parameter vector is adopted if and only if it yields an improvement in the measured energy.
The energy convergence plots with respect to the SPSA iterations, shown in Figure~\ref{fig:energy-flow}, reflect the underlying difficulty of the classical optimization task. The energy for the 2-flow and 4-flow problems shows a relatively smooth convergence towards a low-energy state. In contrast, the energy for the 3-flow problem exhibits significant fluctuations throughout the optimization. This behavior is characteristic of a stochastic optimizer such as SPSA struggling to navigate a rugged energy landscape with many steep, narrow valleys, further supporting the conclusion that this particular 3-flow instance represents a harder optimization problem. 

\vspace{-5pt}
\subsection{QAOA with More Layers}
\begin{figure}[t]
    \centering
    \begin{subfigure}[t]{0.24\textwidth}
        \centering
        \includegraphics[width=\linewidth]{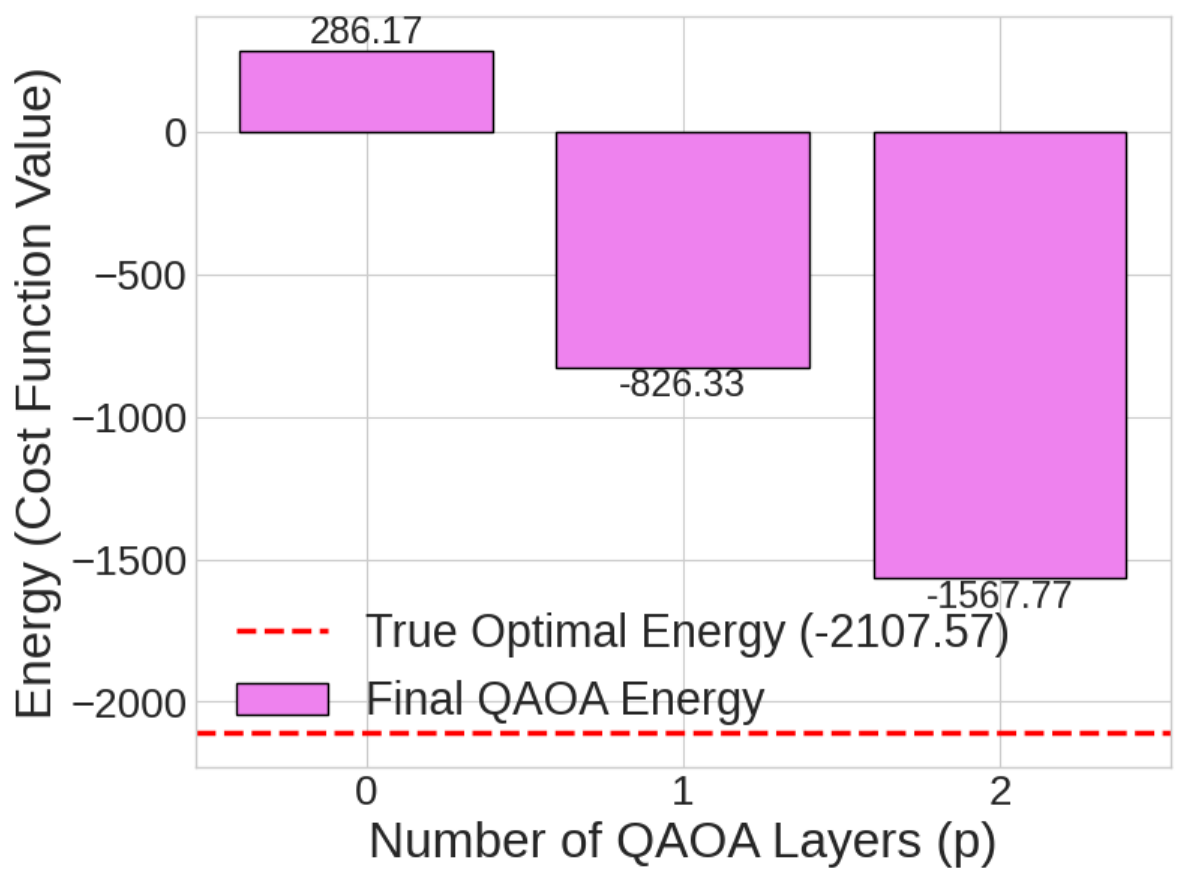}
        \caption{Energy vs. QAOA layers}
        \label{fig:qaoa-energy-vs-layer}
    \end{subfigure}
    \hfill
    \begin{subfigure}[t]{0.24\textwidth}
        \centering
        \includegraphics[width=\linewidth]{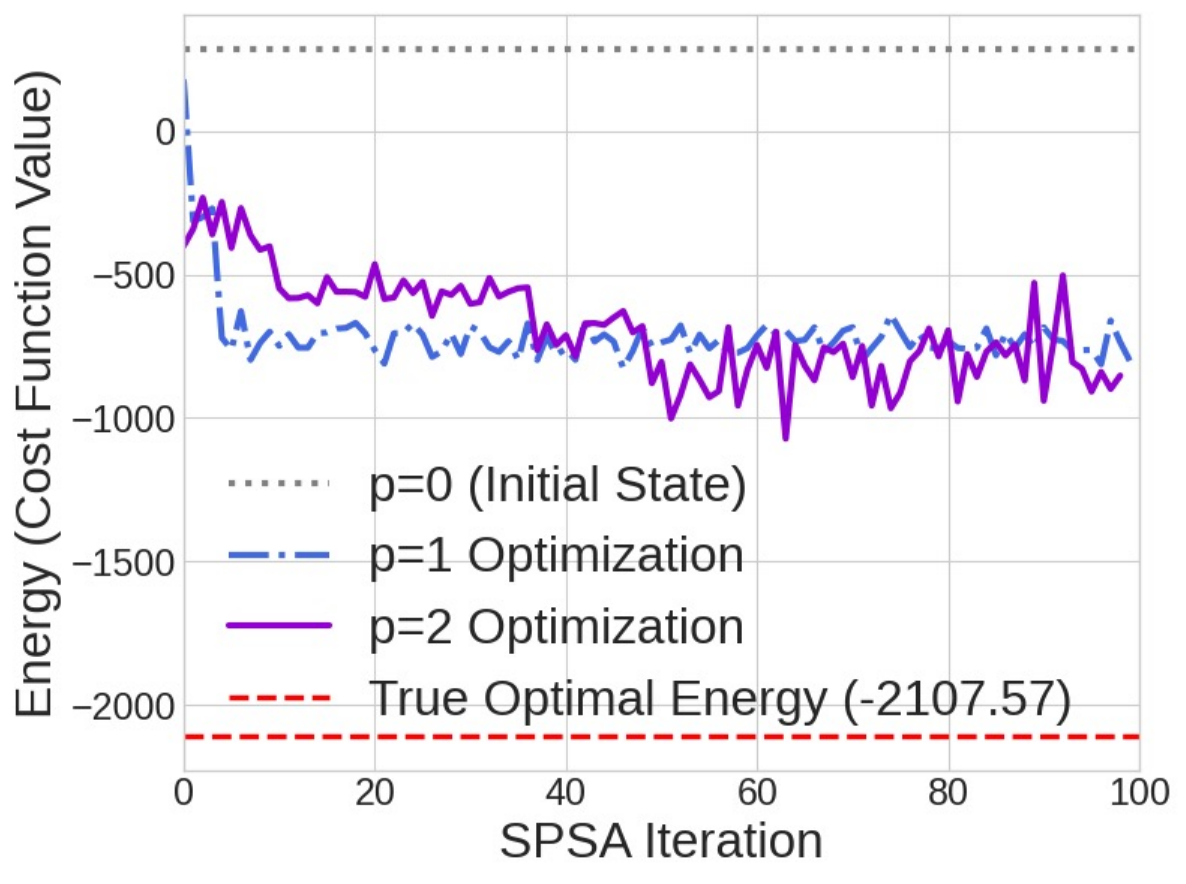}
        \caption{Energy vs. SPSA iterations for different QAOA layers}
        \label{fig:qaoa-energy-vs-layer-2}
    \end{subfigure}
    \caption{The impact of QAOA layers.}
    \vspace{-15pt}
    \label{fig:qaoa-energy}
\end{figure}

\vspace{-5pt}
In an attempt to improve upon the baseline, we tested a deeper, 3-layer (p=3) QAOA. Figure~\ref{fig:qaoa-energy} shows the energy results. Counter-intuitively, this approach failed to yield meaningful results and generally performed worse than the p=1 ansatz. 
This counter-intuitive degradation might be caused by the following two obstacles in variational quantum algorithms. First, each additional QAOA layer increases circuit depth and exposure to gate errors and decoherence. For our device, the cumulative qubit error rate might be 
 large enough to obscure the energy landscape and bias the optimizer toward sub-optimal regions.
Second, deeper QAOA layers expand the parameter space, but the corresponding gradients often vanish exponentially with qubit count, a phenomenon known as \emph{barren plateaus}. Classical optimizers then struggle to locate descent directions, which leads to stagnation around random energy values.

Taken together, the hardware noise and barren-plateau effects might outweigh the theoretical expressiveness advantage of a deeper circuit in our experiments. Therefore, on current NISQ processors, shallow QAOA depths, especially when warm-started, could outperform deeper variants that exceed the machine's coherence budget.

\vspace{-3pt}
\subsection{Computational Resource Usage}
\vspace{-3pt}

The end-to-end runtime for a single, 200-iteration SPSA optimization on the IBM Quantum cloud platform was consistently between 45 and 60 minutes. This runtime is dominated by the overhead of the hybrid quantum-classical loop, which includes circuit transpilation, job submission, queueing, and classical processing of the results at each   function evaluation. This is expected to improve with advancements in both quantum hardware and the surrounding software ecosystem.



%% file: discussion.tex
\vspace{-5pt}
\section{Discussions} \label{sec:discussions}
\vspace{-5pt}

\subsection{Findings}
Through experiments, we demonstrate that the MB-TFSA satellite scheduling problem can indeed be represented effectively as a QUBO, and that current quantum solvers can handle problem instances of small sizes. In particular, we see that the QUBO solutions obtained (via QAOA) are of high quality, confirming that this approach is viable for constellation-scale scheduling. This would be a proof-of-concept that quantum optimization methods can tackle real satellite scheduling tasks, extending prior studies (such as small imaging scheduling problems) to more complex multi-beam communications. 

\vspace{-3pt}
\subsection{Practical Implications}
\vspace{-3pt}

The feasibility of generating high-quality schedules in (near) real-time using a compact quantum/hybrid solver suggests that future satellites or constellations could integrate such technology for autonomous resource management. In a practical scenario, a satellite could  have access to a quantum computing service from the ground in a low-latency link to continuously solve scheduling QUBOs on the fly. Our work would demonstrate a pathway to achieve the responsiveness needed for dynamic traffic where ground-based planning is too slow. For example, if a sudden surge of demand occurs in one beam, the on-board scheduler could quickly re-compute an optimized schedule for the next second of frames, something that classical embedded CPUs might not manage if the problem is very complex. This could improve QoS (higher throughput, lower packet drops) for satellite internet users. In summary, we provide a proof-of-concept that quantum-accelerated decision-making in orbit is plausible in the near future.

%% file: conclusion.tex
\vspace{-5pt}
\section{Conclusion and Outlook}
\vspace{-5pt}

This paper applied quantum computing for aerospace, addressing 
 a real problem of flow scheduling in satellite systems. The study has the potential to accelerate the adoption of hybrid quantum techniques for resource allocation challenges in both terrestrial and space-based communication networks. 
We envision that as quantum hardware progresses (more qubits, less noise), the advantage of such approaches will only grow, allowing larger portions of the scheduling (or similar optimization) problem to be offloaded to quantum solvers. Our work thus serves as a foundation and a case study for the potential of quantum computing in the field of satellite communications and operations, providing guidance and confidence to engineers who might incorporate these technologies in the coming decade.